\documentclass[a4paper]{jpconf}
\usepackage[english]{babel}
\usepackage{graphicx}
\usepackage{amsmath,amssymb}
\usepackage{hyperref} 

\begin{document}
\title{Peres Lattices and chaos in the Dicke model}
\author{Miguel Angel Bastarrachea-Magnani, Jorge G. Hirsch}
\address{Instituto de Ciencias Nucleares, Universidad Nacional Aut\'onoma de M\'exico \\ Apdo. Postal 70-543, M\'exico D. F., C.P. 04510}
\ead{miguel.bastarrachea@nucleares.unam.mx}


\begin{abstract}
Peres lattices are employed as a visual method to identify the presence of chaos in different regions of the energy spectra in the Dicke model.
The coexistence of regular and chaotic regions can be clearly observed for certain energy regions, even if the coupling constant is smaller than the critical value to reach superradiance. It also exhibits the presence of two Excited-State Quantum Phase Transitions, a {\em static} and a {\em dynamic} one. The diagonalization is performed in a extended bosonic coherent basis which enable us to reach a large number of excited states with good numerical convergence.
\end{abstract}


\section{Introduction}

Atom-Field systems have attracted attention from researchers for decades. The Dicke model describes the interaction between $\mathcal{N}$ two-level atoms and a single mode of electromagnetic field \cite{Dicke54}, and it is reach enough the present features as a Quantum Phase Transition at zero temperature, form a normal to a superradiant phase \cite{Hepp73,Wang73}, quantum chaos \cite{Emary03}, Excited-State Quantum Phase Transitions \cite{Per11A}, to name a few. The experimental realization of a Dicke-like Hamiltonian where the superradiant phase was observed for the first time  \cite{Bau10} renewed the interest in the model. The manipulation of systems with just a few interacting qbits connects it with the field of quantum computing and many other interesting potential applications \cite{Garr11}.  

The Dicke Hamiltonian can be written (with $\hbar=1$) as,
\begin{equation}
\begin{split}
H&=\omega a^{\dagger}a+\omega_{0}J_{z}+ \frac{2 \gamma}{\sqrt{\mathcal{N}}}\left( a+a^{\dagger} \right) J_{x},
\end{split}
\end{equation}
where $\omega$ is the field frequency, $\omega_{0}$ the atomic energy level difference, $\gamma$ the coupling, and $J_{x}$, $J_{y}$, and $J_{z}$ are collective pseudo-spin atomic operators which obey the SU(2) algebra. In what follows we restrict the analysis to the atomic symmetric subspace with
 $j=\mathcal{N}/2$ (the pseudo-spin lenght), being $j(j+1)$ the eigenvalue of $\mathbf{J}^{2}=J_{x}^{2}+J_{y}^{2}+J_{z}^{2}$.
 It has a ground state QPT whose critical coupling is (in the thermodynamical limit $\mathcal{N}\rightarrow 0$) $\gamma_{c}=\sqrt{\omega_{0}\omega}/2$.  

The Dicke model has an integrable approximation (employing the rotating wave approximation) called the Tavis-Cummings (TC) model \cite{TC68}. 
The integrability of the TC model, comes from the conserved quantity $\Lambda$, the total number of excitations \cite{Cas09}:
\begin{equation}
\Lambda=a^{\dagger}a+J_z+j,
\end{equation}
with eigenvalues $\lambda = n + j + m$, where $n$ is the number of photons, $m$ the eigenvalue of $J_z$ and $j+m$ the number of excited atoms. 
The presence of the counter-rotating terms in the Dicke Hamiltonian prevents it to commute with $\Lambda$. It does commute, however, with a parity operator $\Pi$:
\begin{equation} \label{parity}
\Pi=exp(i\pi\Lambda) .
\end{equation}
It allows to separate the eigenstates into those with positive and negative parity, but it is not enough to classify all states, implying that
Dicke Hamiltonian is not integrable. 
The Dicke model becomes integrable in two limits of the Hamiltonian parameters, the zero-coupling limit $\gamma\rightarrow 0$, and the limit with $\omega_{0}\rightarrow 0$ \cite{Basta11}. As a non-integrable Hamiltonian, there are not an analytical solutions, except for asymptotic solutions which are exact for the ground state in the thermodynamical limit \cite{Emary03,Cas11}. 

Cavity QED applications \cite{Fin09} are described by the Dicke Hamiltonian with a limited number of qbits, whose eigenstates can only be obtained employing numerical methods. 
As the bosonic sector is infinite dimensional, a truncation $N_{max}$ in the number of bosons is necessary. 
To perform the numerical diagonalization of the Hamiltonian we use an extended bosonic coherent basis \cite{Basta11,Basta12}. This basis enables us to obtain a large fraction of converged states employing a single value of the truncation $N_{max}$ . An upper bound for the numerical precision in the wave function for every state can be assigned \cite{Basta13}, as briefly reviewed in Appendix 1. With this tool it is possible to explore a relevant sector of the energy spectrum for relatively large values to the number of atoms $\mathcal{N}$, in the superradiant phase.     

A very interesting feature present in some quantum spectra, the Excited-State Quantum Phase Transition (ESQPT) has been found in the Dicke and TC models \cite{Per11A}. An ESQPT takes place along the energy spectrum, for fixed values of the Hamiltonian parameters. It is manifested by singularities in the level density, order parameters, and wave function properties \cite{Cap08}. An ESQPT can have important effects in decoherence \cite{Rel08} and the temporal evolution of quantum quenches \cite{Per11B}. Some average properties of the observables, including the ESQPT, have been analyzed recently in the thermodynamic limit \cite{Bran13}. 

The purpose of this work is to explore the presence of regular and chaotic regions along the energy spectra employing Peres lattices in the Dicke model. In section 2 we describe Peres method and the Peres lattice idea. In section 3 we show and discuss the Dicke Hamiltonian Peres lattices. We employ them to characterize the non-integrability of Dicke model and to study qualitatively its spectrum. Finally, in section 4 we show the conclusions. 

\section{Peres lattices and integrability}

The notion of integrability comes from classical physics. A classical integrable system has as many conserved quantities as degrees of freedom, so it can be solved analytically. Its orbits in phase space lie on tori, and if we employ a Poincar\'e section, the phase space trajectories can be seen as regular curves. The non-integrability in a classical system is strongly bonded to chaos. If we add a perturbation term to the classical Hamiltonian, the conserved quantities could stop to be so and in that case the system becomes non-integrable. Also, as the strength of perturbation increases the tori get destroyed and the regular orbits become chaotic ones. Again, this can be seen easily using a Poincar\'e section, irregular patterns appear where regular orbits were before. Then, with non-integrability chaos arises \cite{Sal98,Gold02}. For a quantum system, however, the definition of integrability cannot be the same as in classical physics, mainly because there are no trajectories in phase space where chaotic and regular orbits can be distinguished, and there is not a direct analog of the Poincar\'e sections that help us to identify the integrability of a Hamiltonian and the presence of chaos. 

A quantum system whose classical analog is integrable must have a ``regular'' spectrum, i. e. its energy levels can be labeled in a natural way by quantum numbers related to the constant of motions. Following this idea, A. Peres proposed a visual method that plays a role similar of Poincar\'e sections in classical mechanics. \cite{Per84}, and helps to identify the integrability of a quantum system by its regularity in the spectrum. 
On the other hand, for a non-integrable system the absence of this regularity can be observed. 

When a quantum system is integrable, associated with an unperturbed Hamiltonian $H_{0}$, a plot of the energies of individual states versus the respective eigenvalues of a constant of motion $P$, which satisfies $[H_{0},P]=0$, form a lattice of regularly distributed points. When the system is perturbed, $H=H_{0}+\gamma H_{1}$ and becomes non-integrable, $P$ is not anymore a conserved quantity. However, as Peres pointed out, we can use the expectation values of a $P$ (now called Peres operator) in the energy representation and plot them against the Hamiltonian eigenvalues. Peres noted that, in classical physics, the time-average of any bounded quantity depending in canonical variables is trivially a constant of motion. For quantum systems this would correspond to the expectation values of the Peres operator. 

The choice of $P$ connects the unperturbed and perturbed cases and such plots are called Peres lattices. If the system is integrable the Peres lattice is regular, as each energy level has a natural way to be labeled by the quantum number. A smooth perturbation doesn't destroy the lattice, instead a set of localized distortions is created, and the rest of the lattice remains regular. This can be seen as the coexistence of chaotic an regular behaviors in a classical system with mixed dynamics, i. e. the remnants of the tori. Finally, as the perturbation increases the system becomes non-integrable and is characterized by an irregular lattice. In this way the Peres method represents a qualitatively sensitive probe into the competition between regular and chaotic behaviors in the quantum spectrum of a system \cite{Str09}. Moreover, the freedom in choosing the Peres operator makes it possible to focus on various properties of individual states and to closely follow the way how chaos sets in and proliferates in the system.

Therefore, Peres lattices allow us to observe changes in the spectrum qualitatively. We employ the Peres method to analyze the features of the Dicke Hamiltonians spectrum, the onset of chaos and its ESQPT.

\section{Peres lattices in Dicke model}

As the Dicke Hamiltonian has two degrees of freedom, in order to build the Dicke model Peres lattices we only need a single Peres operator. Thanks to the freedom in the choice of it we can explore different features with different Peres operators. It is worth to be mentioned that, as the Dicke Hamiltonian commutes with the parity operator, it is important to separate states with different parity in order to build the Peres lattices. In appendix 2 we show the extended coherent basis with well defined parity, which we use to diagonalize the Hamiltonian taking advantage of this symmetry. Taking into account this considerations we choose as Peres operators  $J_{z}$, $J_{x}^{2}$ (as $J_{x}$ is not a good operator because connects states of different parity) and $n=a^{\dagger}a$.

In what follows we show several Peres lattices for different values of the coupling constant $\gamma$, in resonance for $\omega = \omega_0 = 1.0$. 
The critical coupling takes the value $\gamma_c = 0.5$. The number of atoms is fixed at $\mathcal{N}=40$.

\begin{figure}
\begin{tabular}{cc}
a) & b) \\
\includegraphics[angle=0,width=0.5\textwidth]{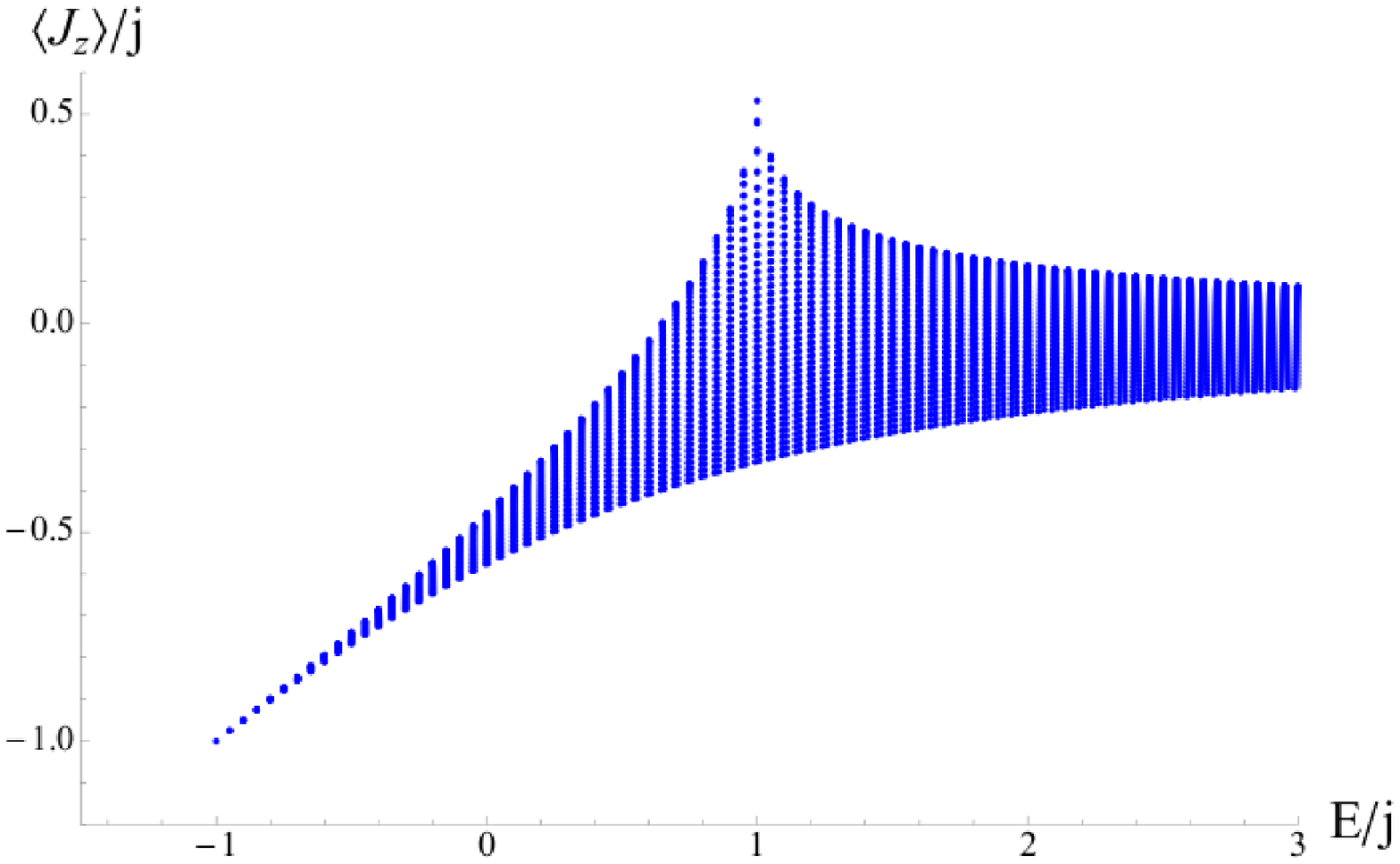} & \includegraphics[angle=0,width=0.5\textwidth]{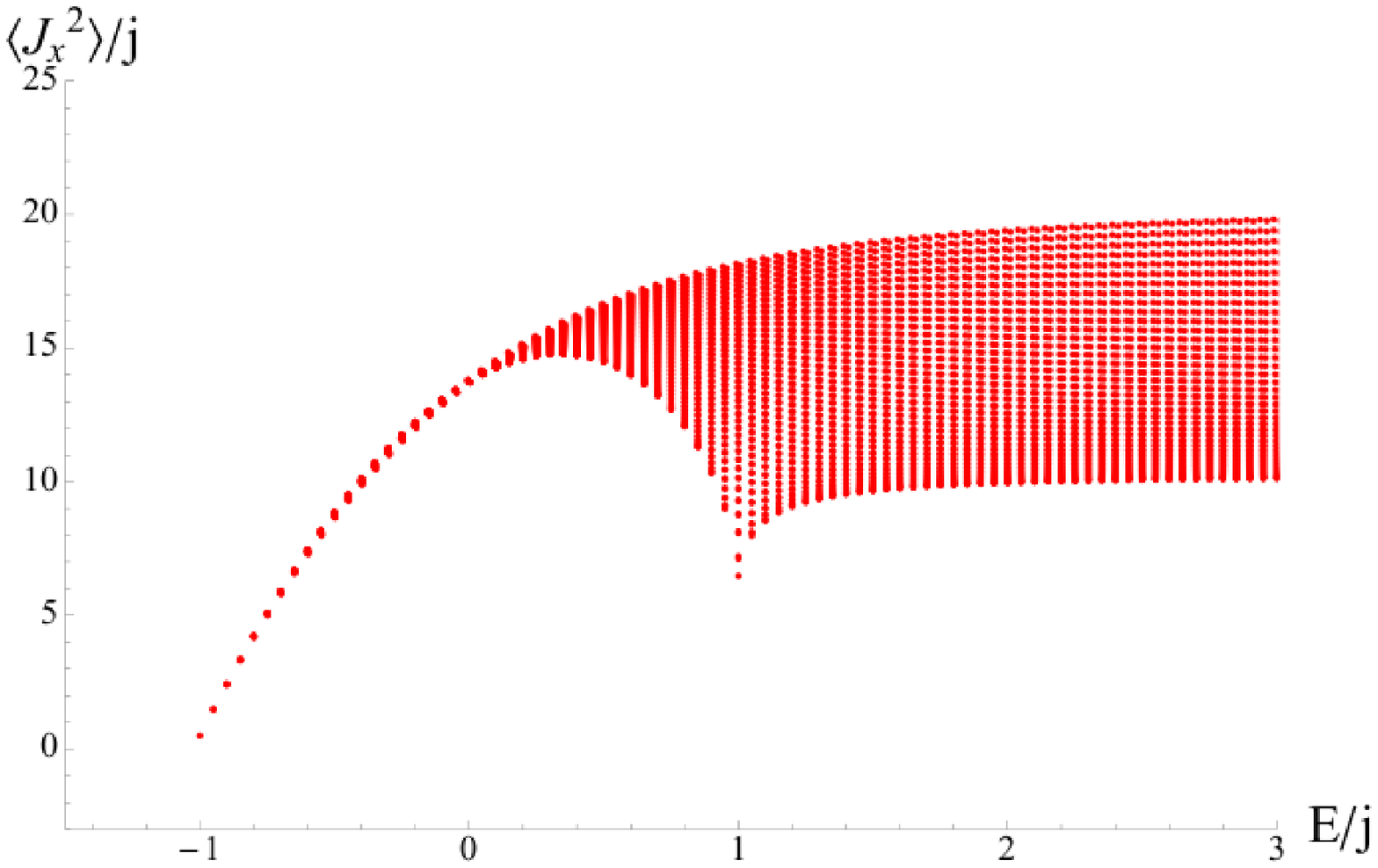} \\
c) & d) \\
\includegraphics[angle=0,width=0.5\textwidth]{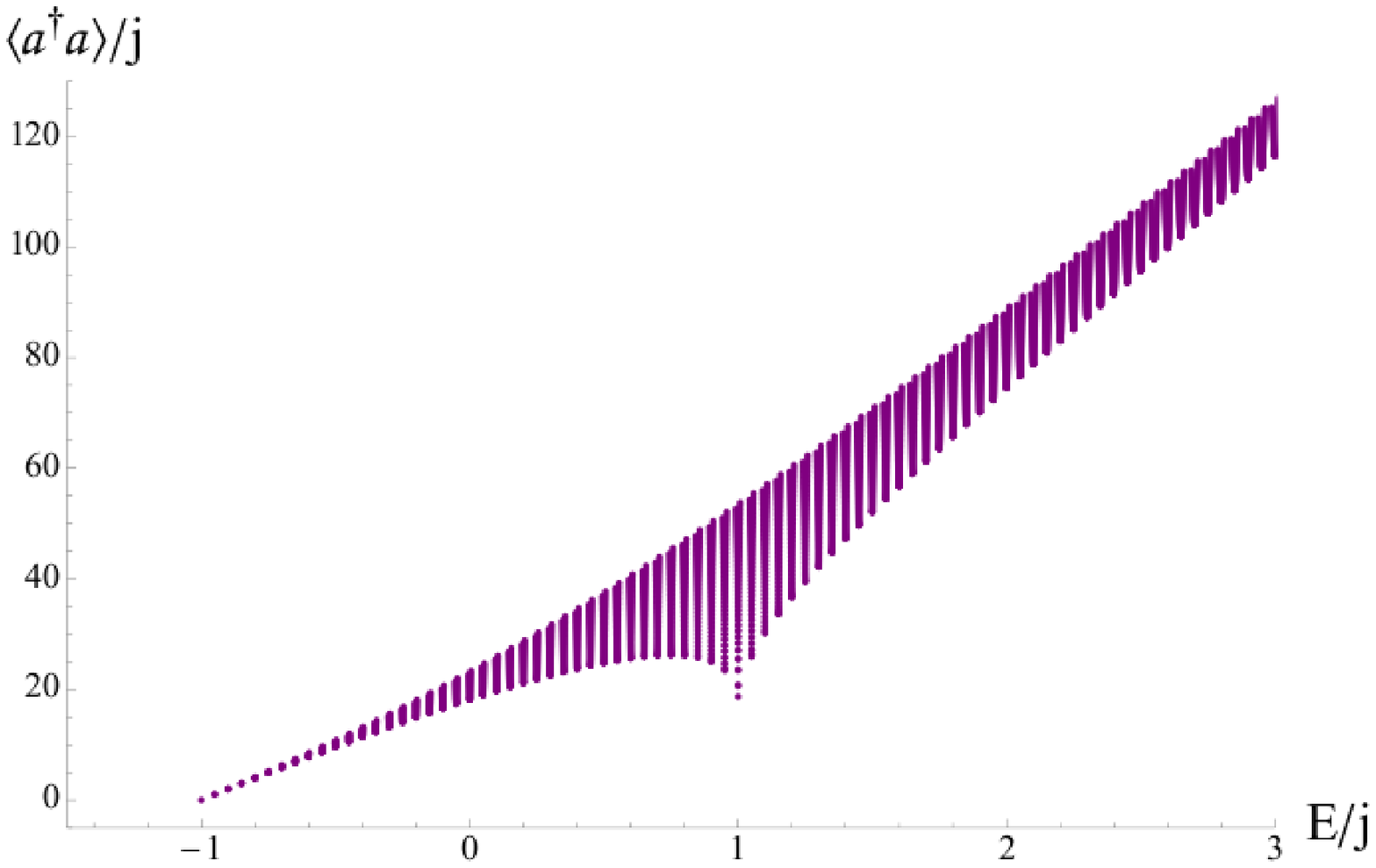} & \includegraphics[angle=0,width=0.5\textwidth]{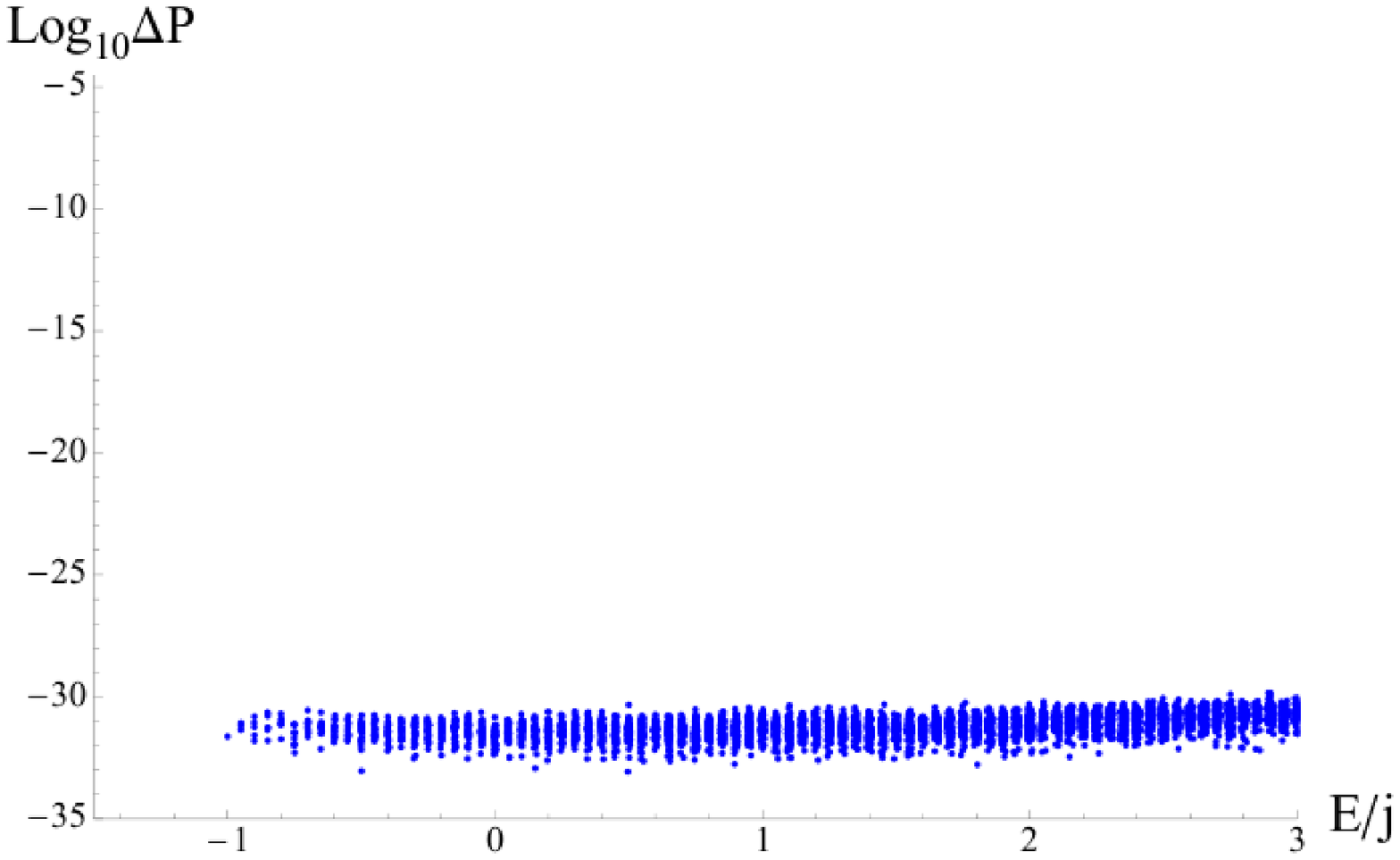} \\
\end{tabular}
\caption{Peres lattices for the Dicke model. Normal phase near zero-coupling $\gamma=0.01\gamma_{c}$, with $N_{mnax}=250$. Peres operators: $J_{z}$ (a), $J_{x}^{2}$ (b) and $n$ (c). The numerical precision in the wave function for each individual state is shown in (d).}  
\label{fig1}
\end{figure}

For values of the coupling close to zero, $\gamma=0.01\gamma_{c}$, we present in figure \ref{fig1} the Peres lattices. As it can be seen from the lattices, they exhibit regular patterns. This is just what we expect as the coupling is near the integrable limit $\gamma \rightarrow 0$. 
In this case the energies are $E(n,j,m) \approx n+ m$. The Hamiltonian is nearly integrable, we can label each state with their quantum numbers, in this case the number of atoms and the number of photons and thus the Peres lattices are regular.  Degenerate states are visualized as vertical lines of dots with the same energy and different expectation values of the Peres operators. The ground state has $n=0$, $m=-j$ and $\frac {E}{j} = -1$. The degeneracy grows linearly with the energy up to $2j+1$, and remains constant beyond this value, which corresponds to $\frac {E}{j} = 1$, which is clearly seen as a well defined peak in the Peres lattices, Figs 1 a, b and c. Fig. 1d shoes the upper bound of the numerical error in each wave function due to the truncation. All of them are smaller that $10^{-30}$.

\begin{figure}
\begin{tabular}{cc}
a) & b) \\
\includegraphics[angle=0,width=0.5\textwidth]{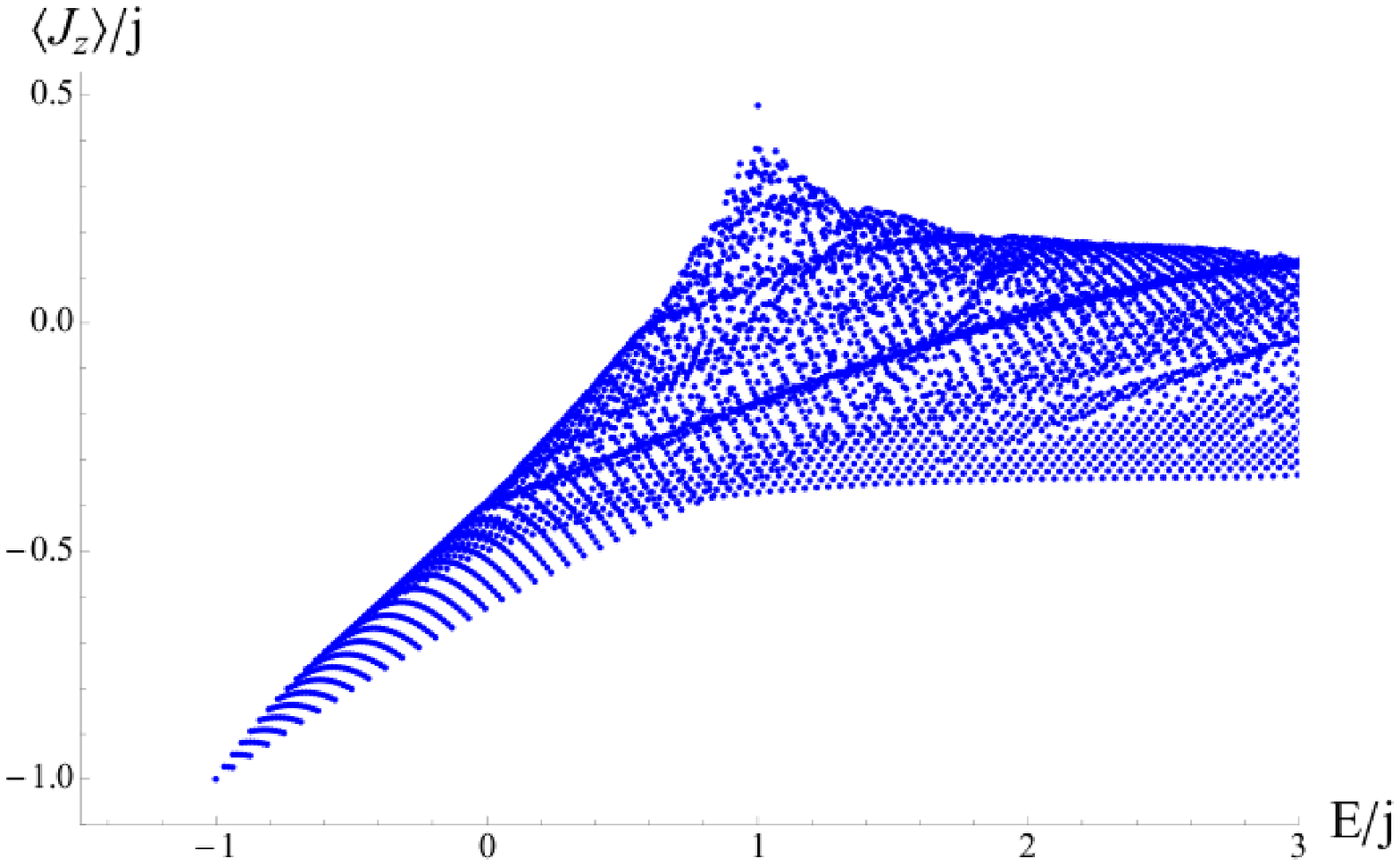} & \includegraphics[angle=0,width=0.5\textwidth]{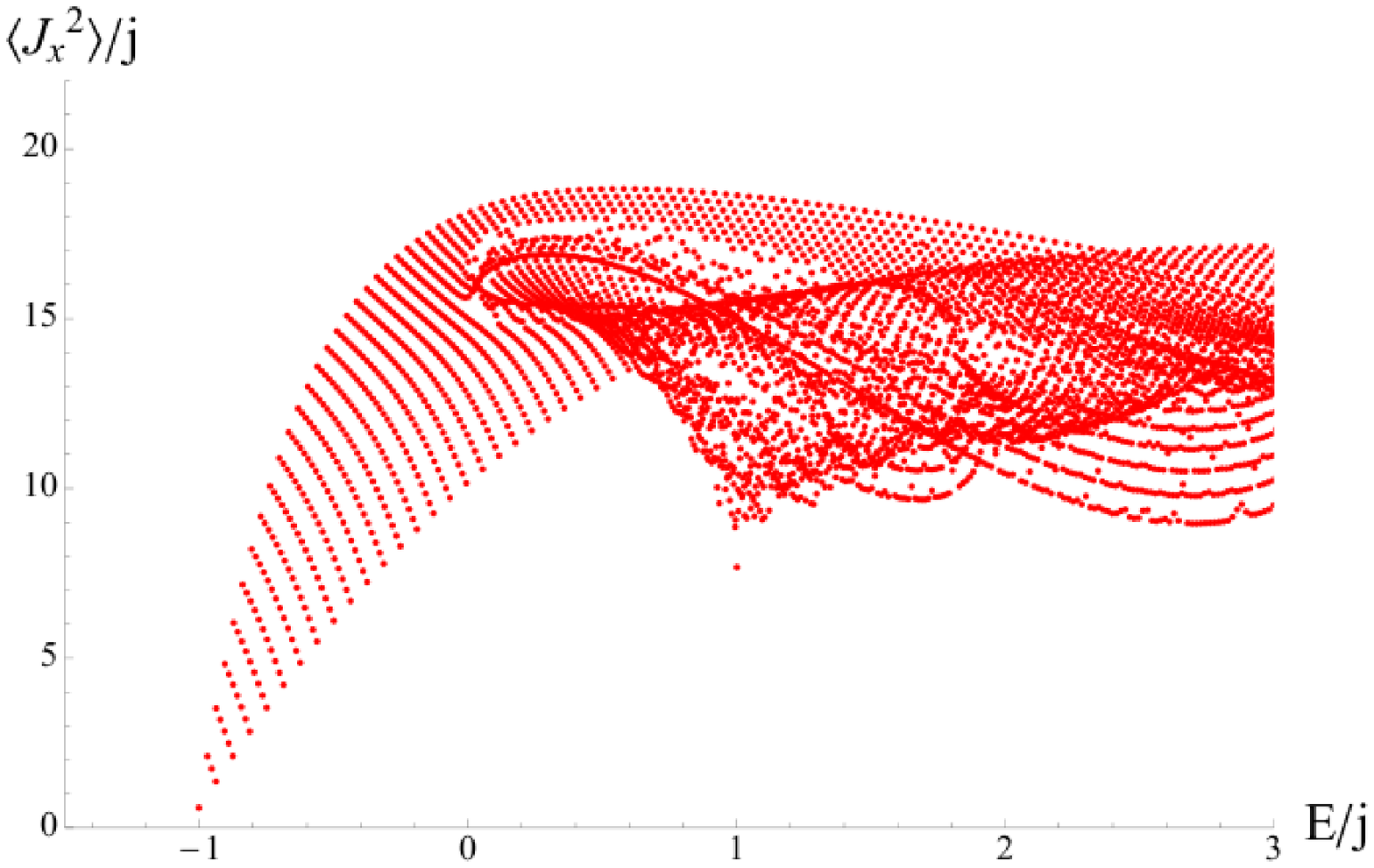} \\
c) & d) \\
\includegraphics[angle=0,width=0.5\textwidth]{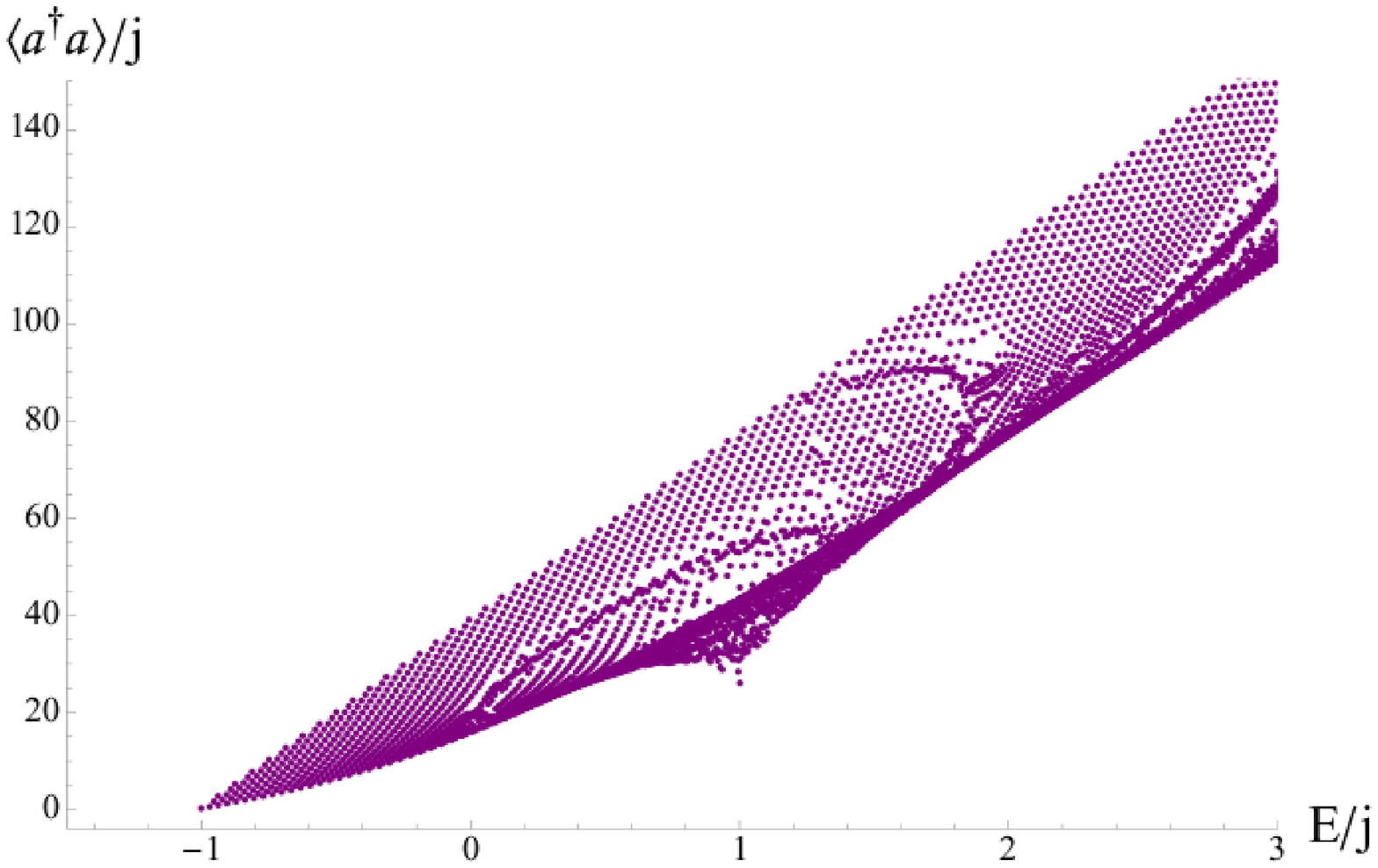} & \includegraphics[angle=0,width=0.5\textwidth]{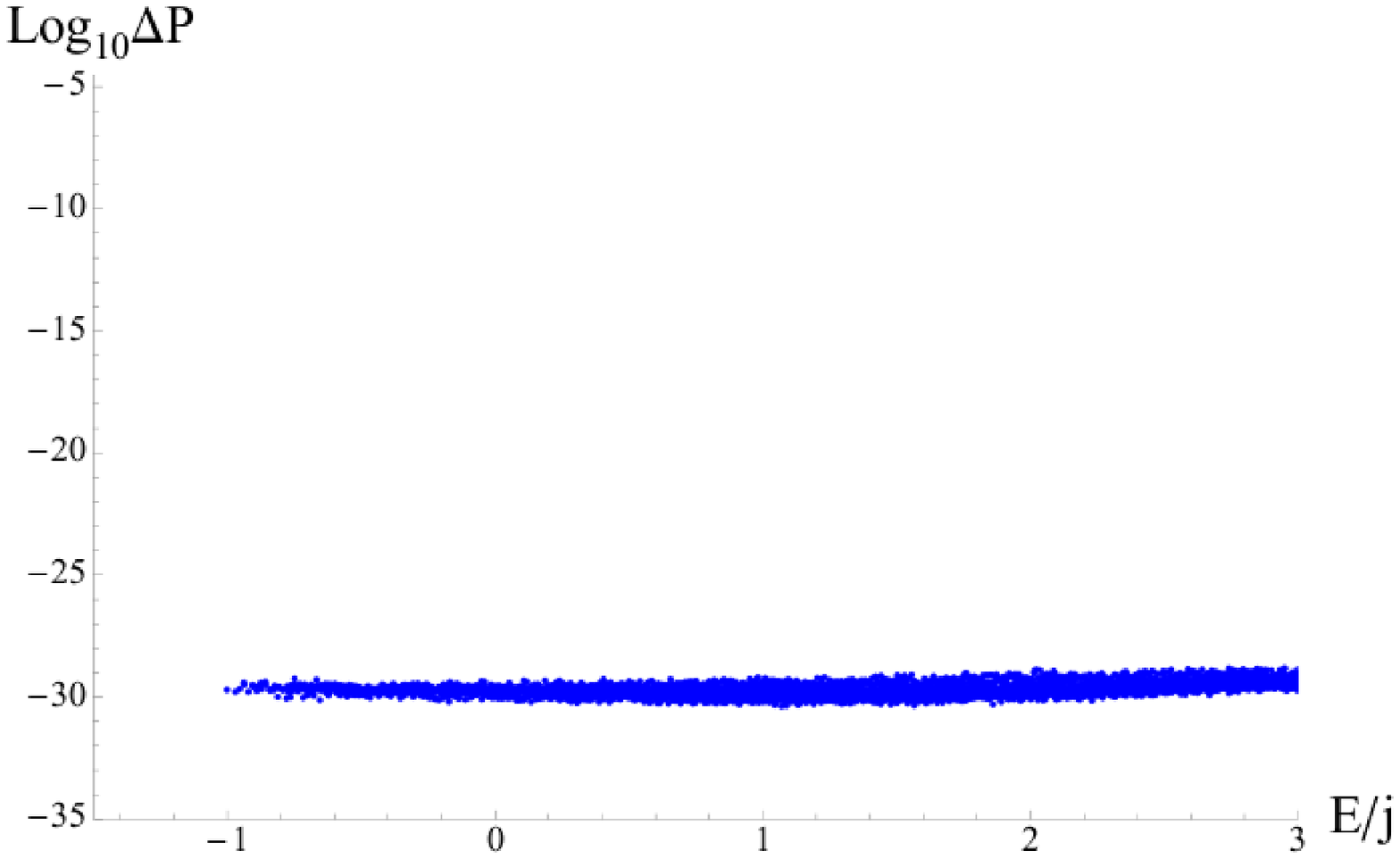} \\
\end{tabular}
\caption{Peres lattices for the Dicke model. Normal phase $\gamma=0.6\gamma_{c}$, with $N_{mnax}=250$. Peres operators: $J_{z}$ (a), $J_{x}^{2}$ (b) and $n$ (c). The numerical precision in the wave function for each individual state is shown in (d).}  
\label{fig2}
\end{figure}

In  figure \ref{fig2} we present the Peres lattices for a small but non negligible coupling $\gamma=0.03\gamma_{c}$. 
Although the system is the regular phase, irregular regions in the energy spectrum can be observed for $E>0$, reflecting the fact that
the Hamiltonian is no longer integrable. We can guarantee that the observed effects are robust, because the numerical noise remains around $10^{-30}$

\begin{figure}
\begin{tabular}{cc}
a) & b) \\
\includegraphics[angle=0,width=0.5\textwidth]{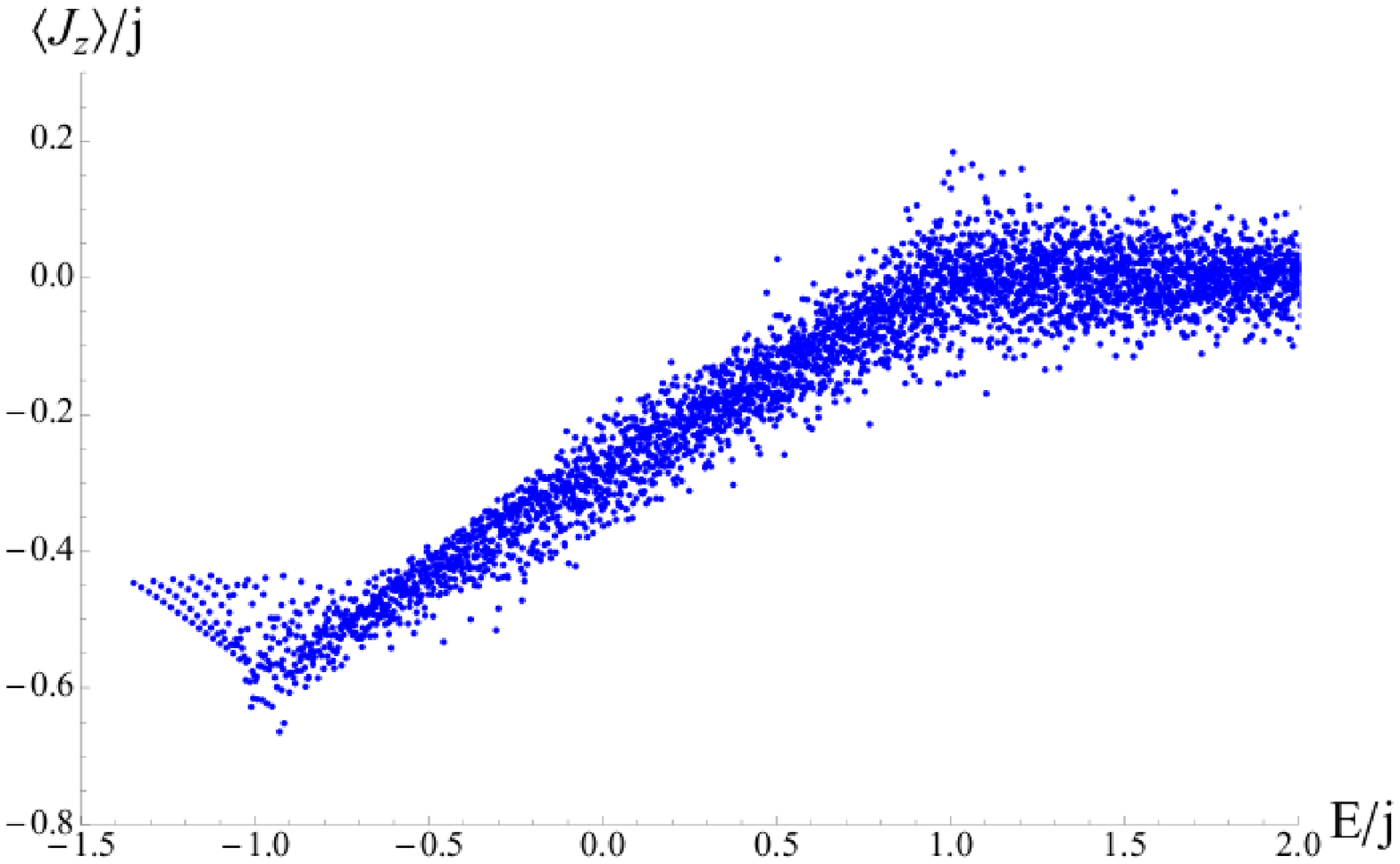} & \includegraphics[angle=0,width=0.5\textwidth]{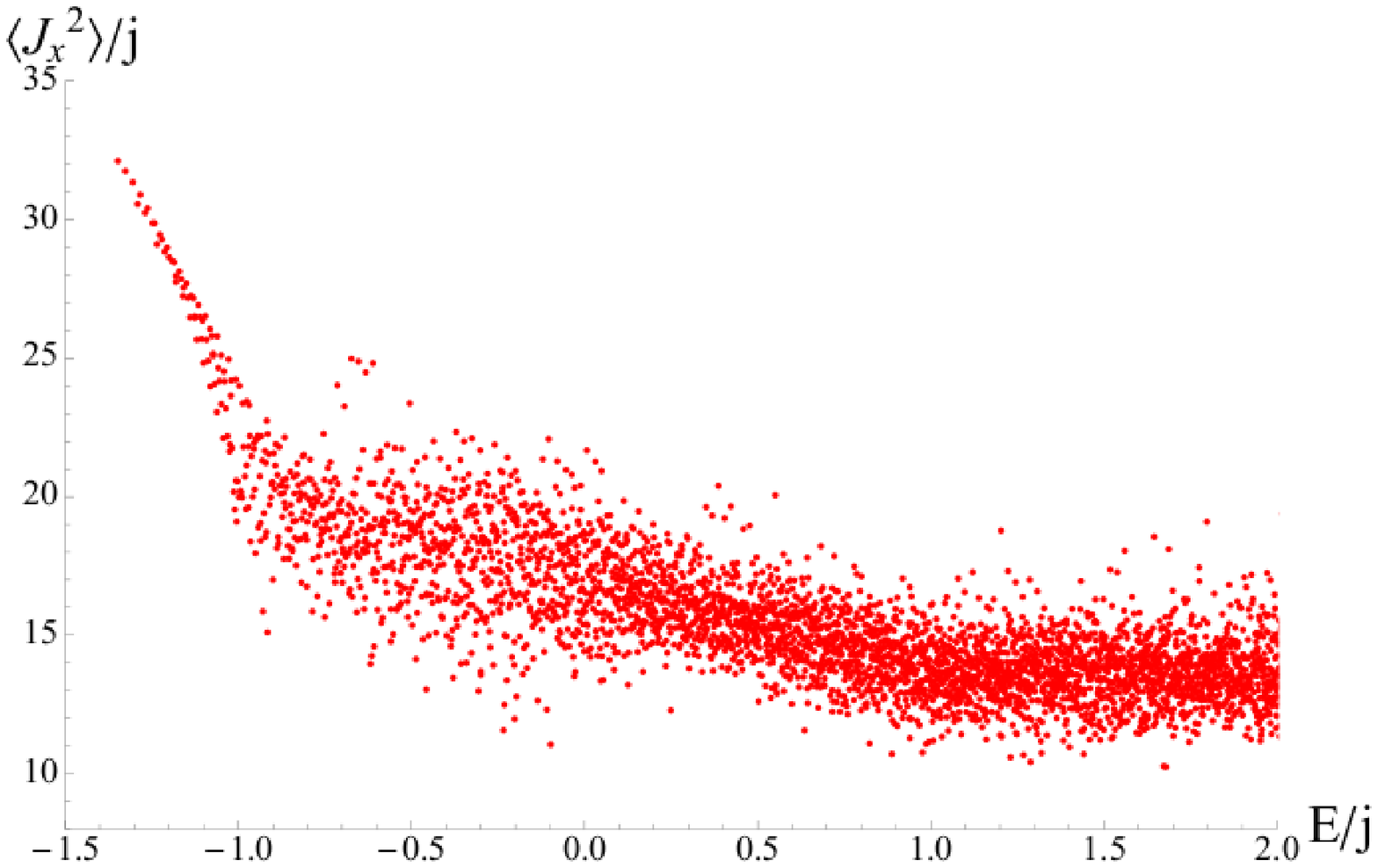}\\
c) & d) \\
\includegraphics[angle=0,width=0.5\textwidth]{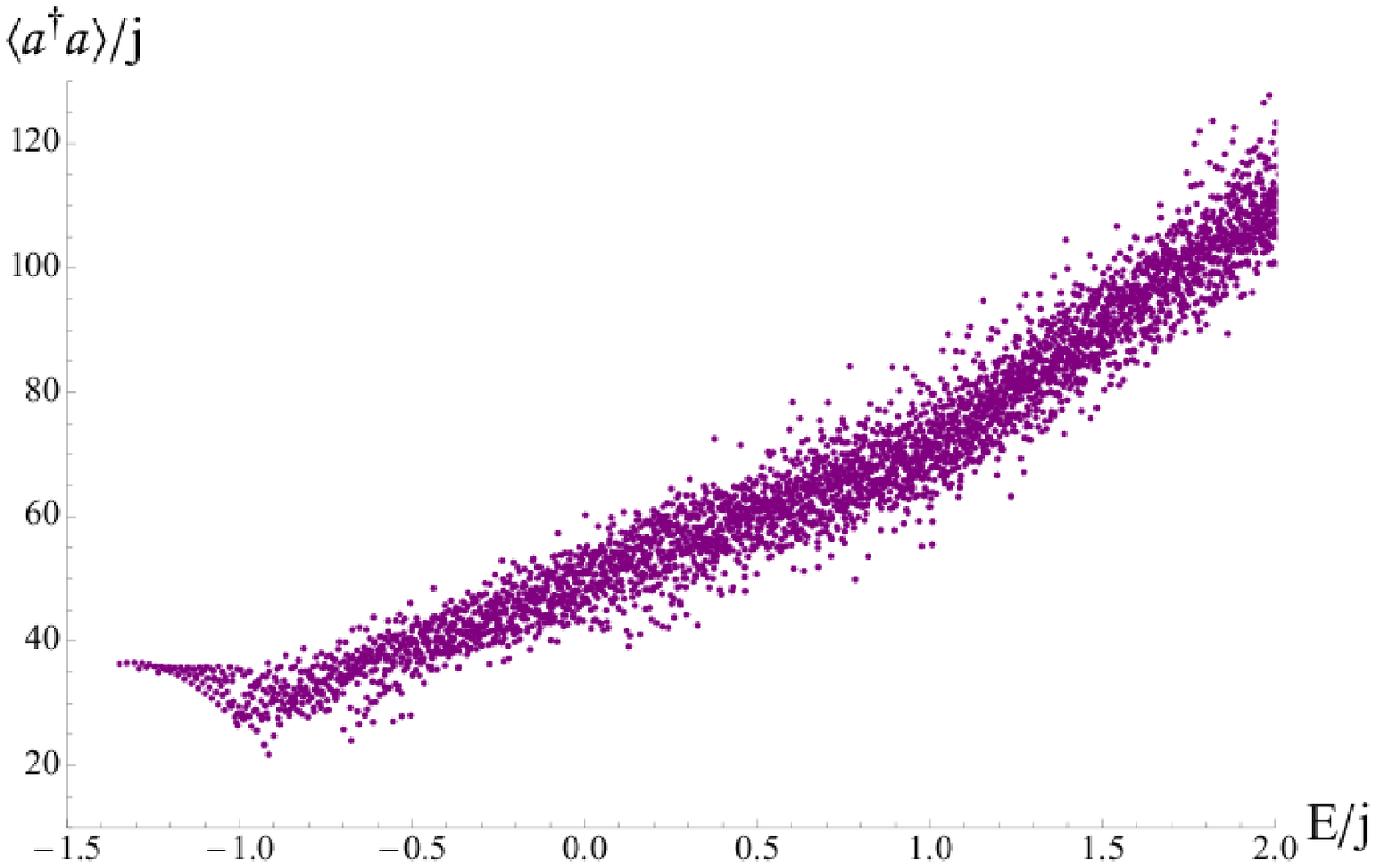} &\includegraphics[angle=0,width=0.5\textwidth]{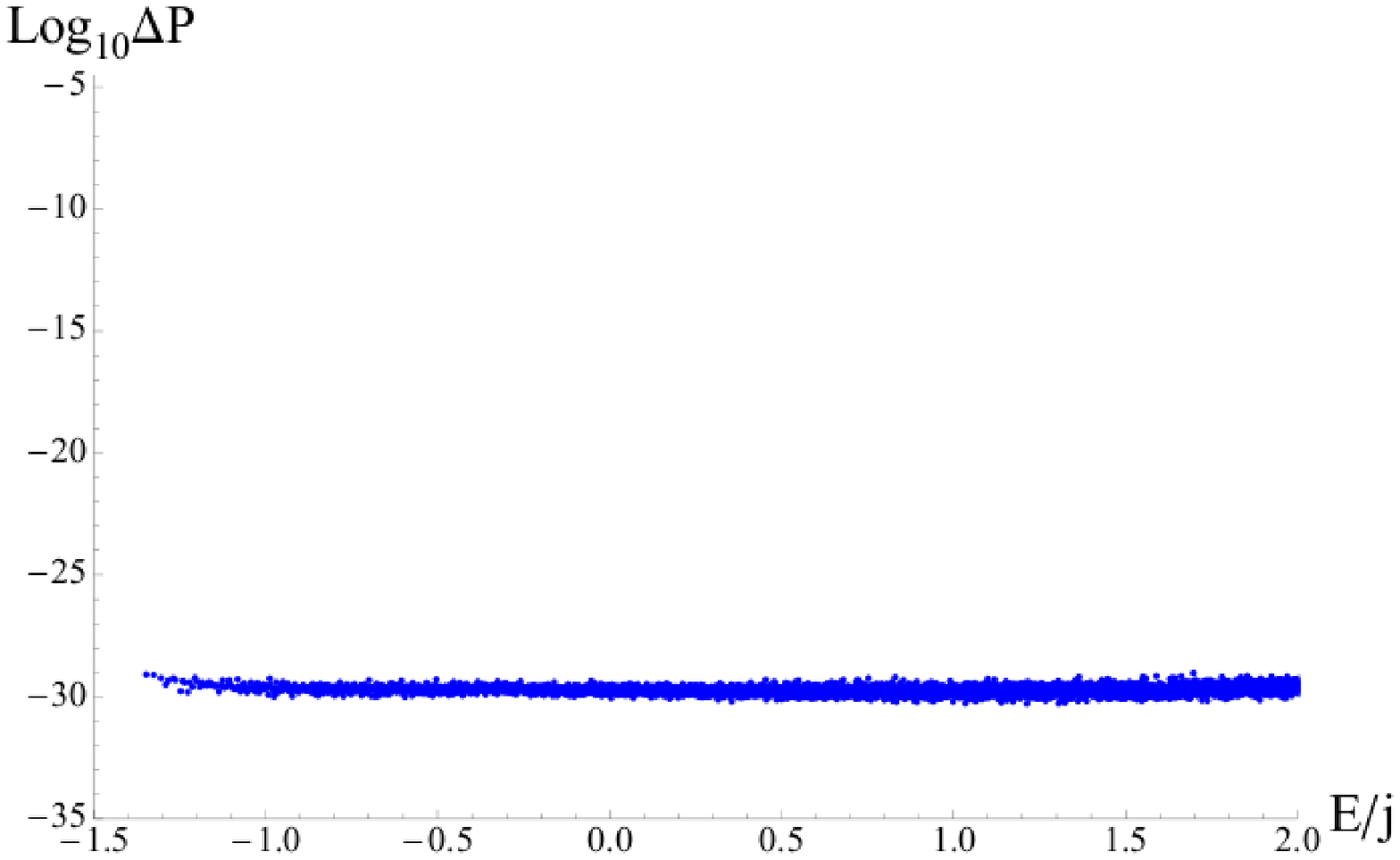} \\
\end{tabular}
\caption{Peres lattices for the Dicke model. Superradiant phase $\gamma=1.5\gamma_{c}$ with $N_{mnax}=250$. Peres operators: $J_{z}$ (a), $J_{x}^{2}$ (b) and $n$ (c). The numerical precision in the wave function for each individual state is shown in (d).}  
\label{fig3}
\end{figure} 

\begin{figure}
\begin{tabular}{cc}
a) & b) \\
\includegraphics[angle=0,width=0.5\textwidth]{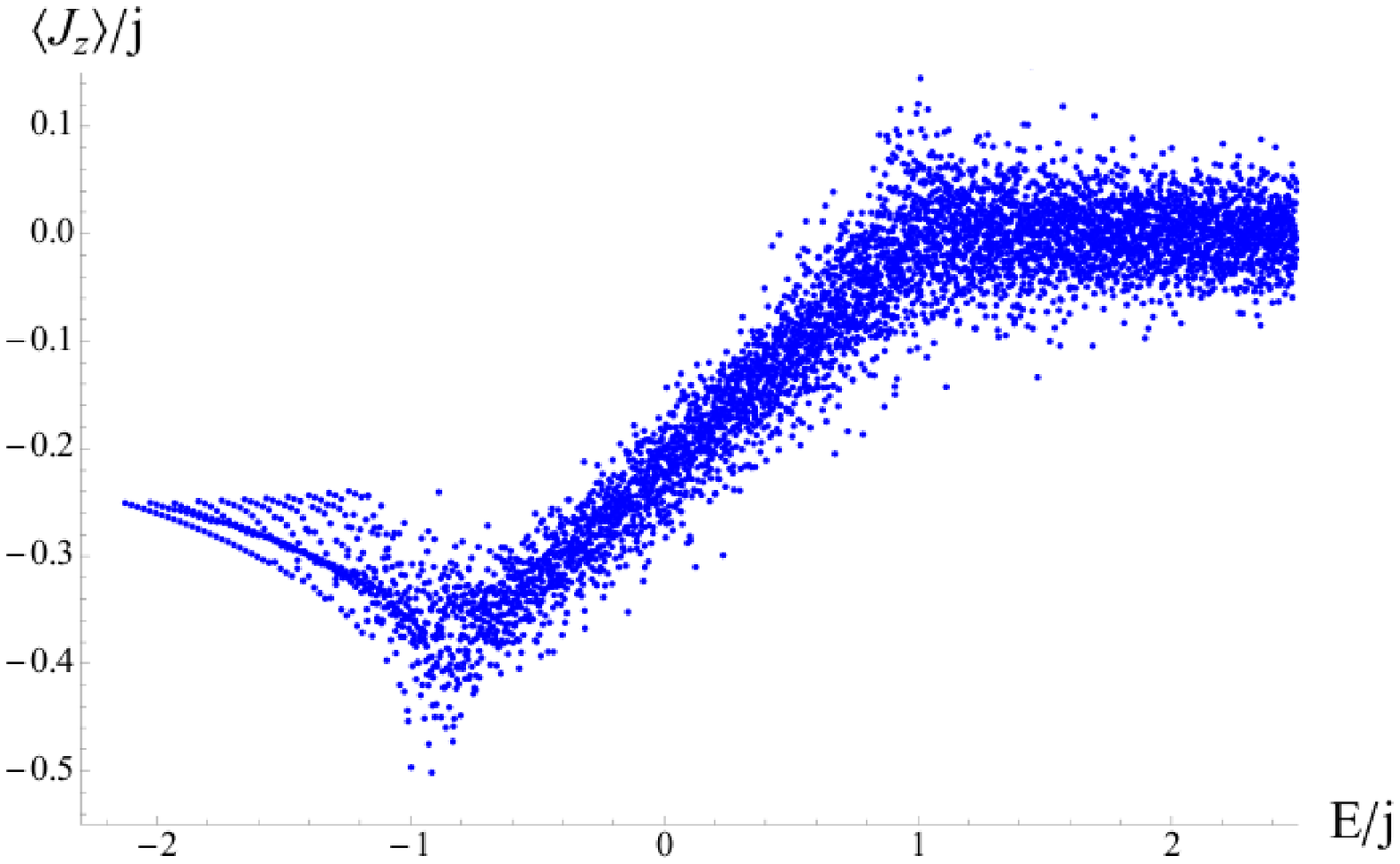} & \includegraphics[angle=0,width=0.5\textwidth]{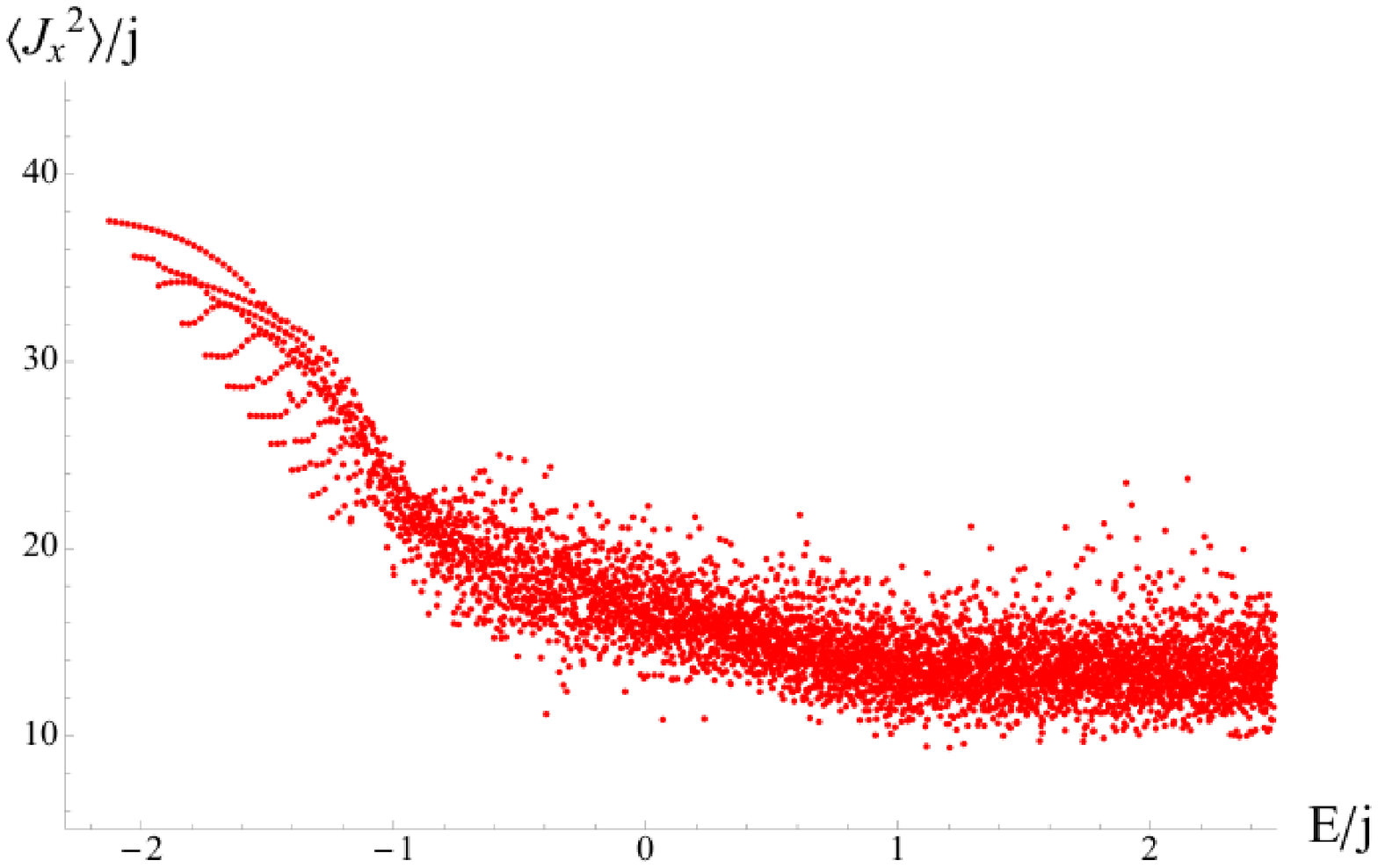}\\
c) & d) \\
\includegraphics[angle=0,width=0.5\textwidth]{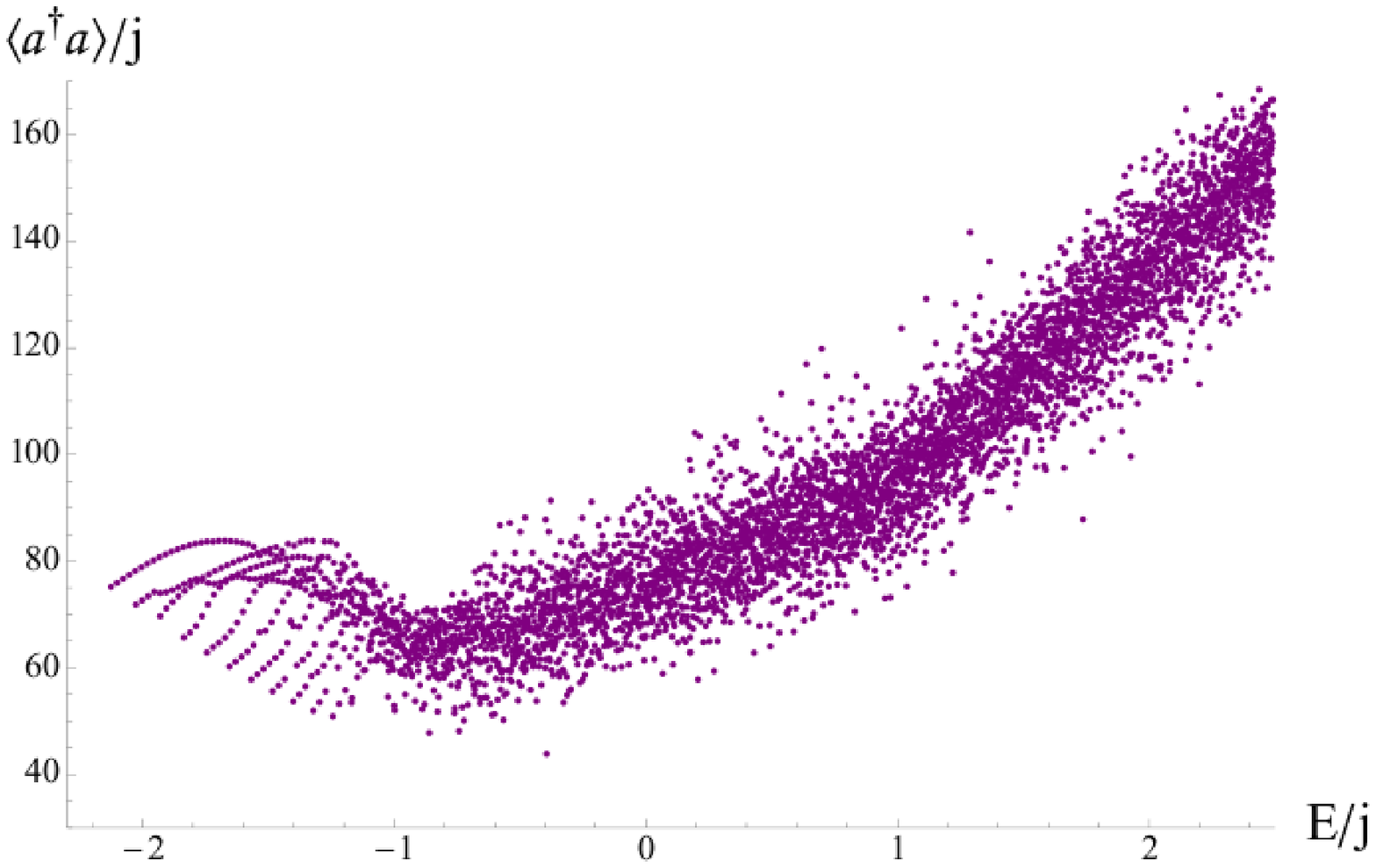} &\includegraphics[angle=0,width=0.5\textwidth]{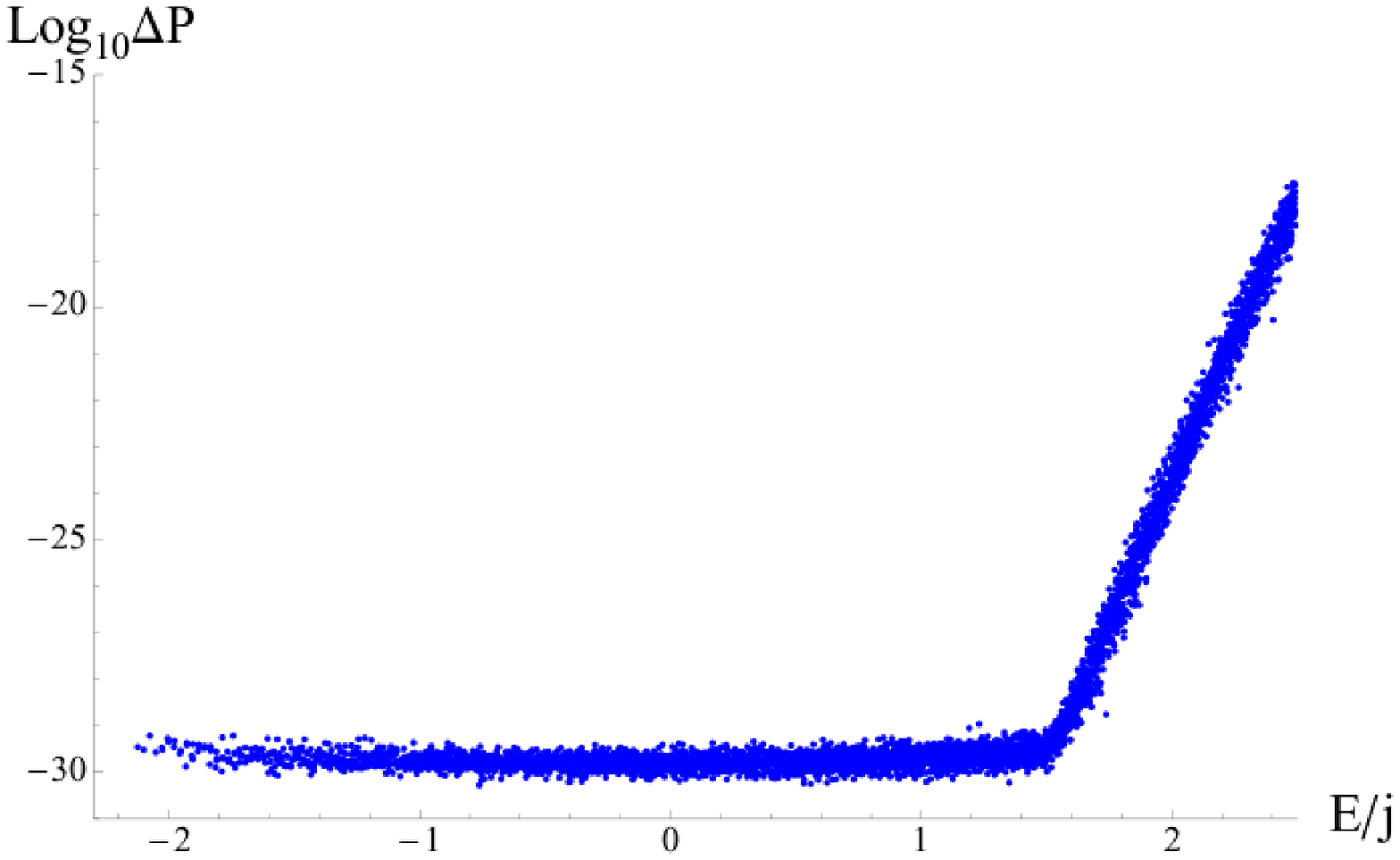} \\
\end{tabular}
\caption{Peres lattices for the Dicke model. Superradiant phase $\gamma=2.0 \gamma_{c}$, with $N_{mnax}=250$. Peres operators: $J_{z}$ (a), $J_{x}^{2}$ (b) and $n$ (c). The numerical precision in the wave function for each individual state is shown in (d).}  
\label{fig4}
\end{figure} 

\begin{figure}
\begin{tabular}{cc}
a) & b) \\
\includegraphics[angle=0,width=0.5\textwidth]{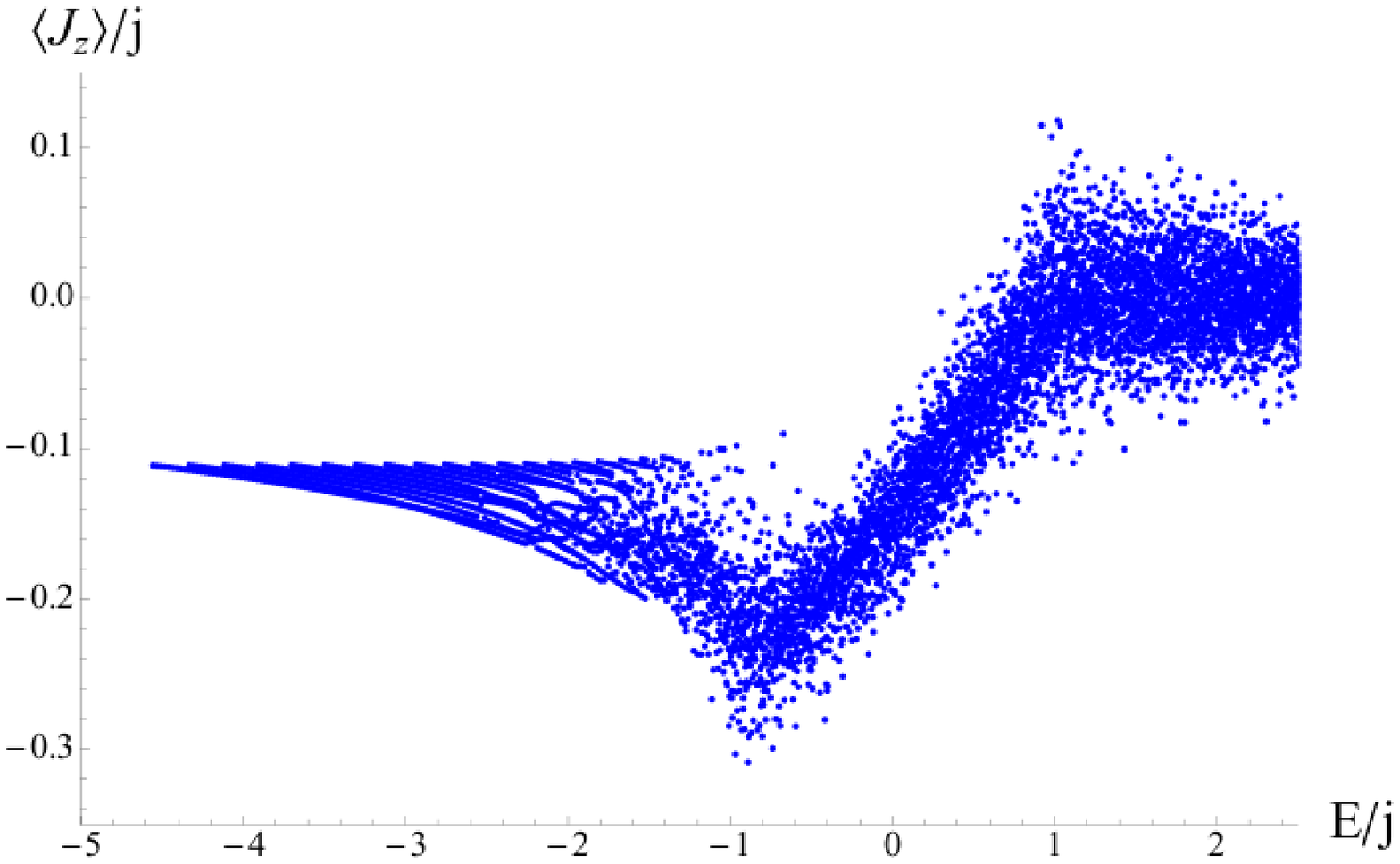} & \includegraphics[angle=0,width=0.5\textwidth]{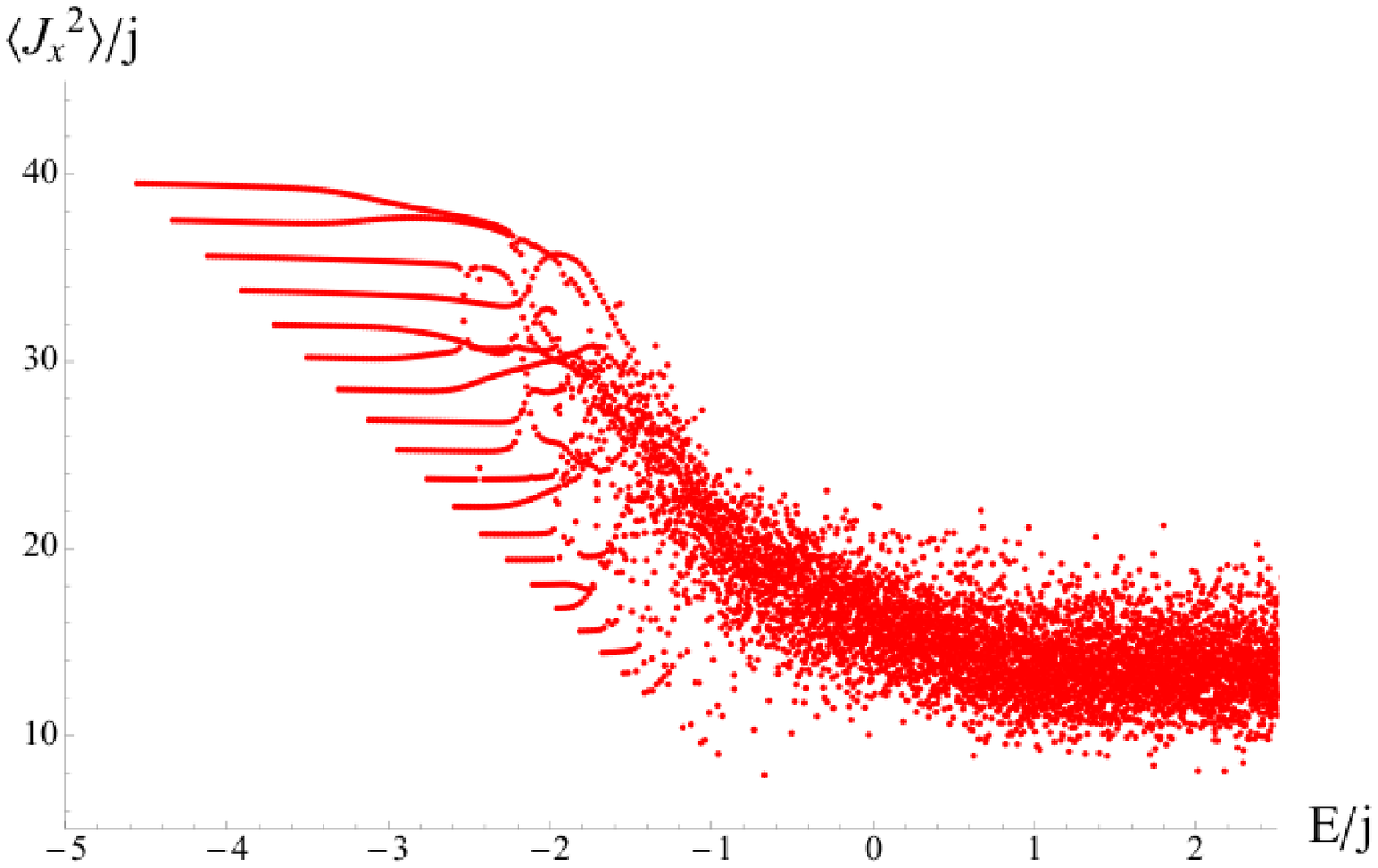}\\
c) & d) \\
\includegraphics[angle=0,width=0.5\textwidth]{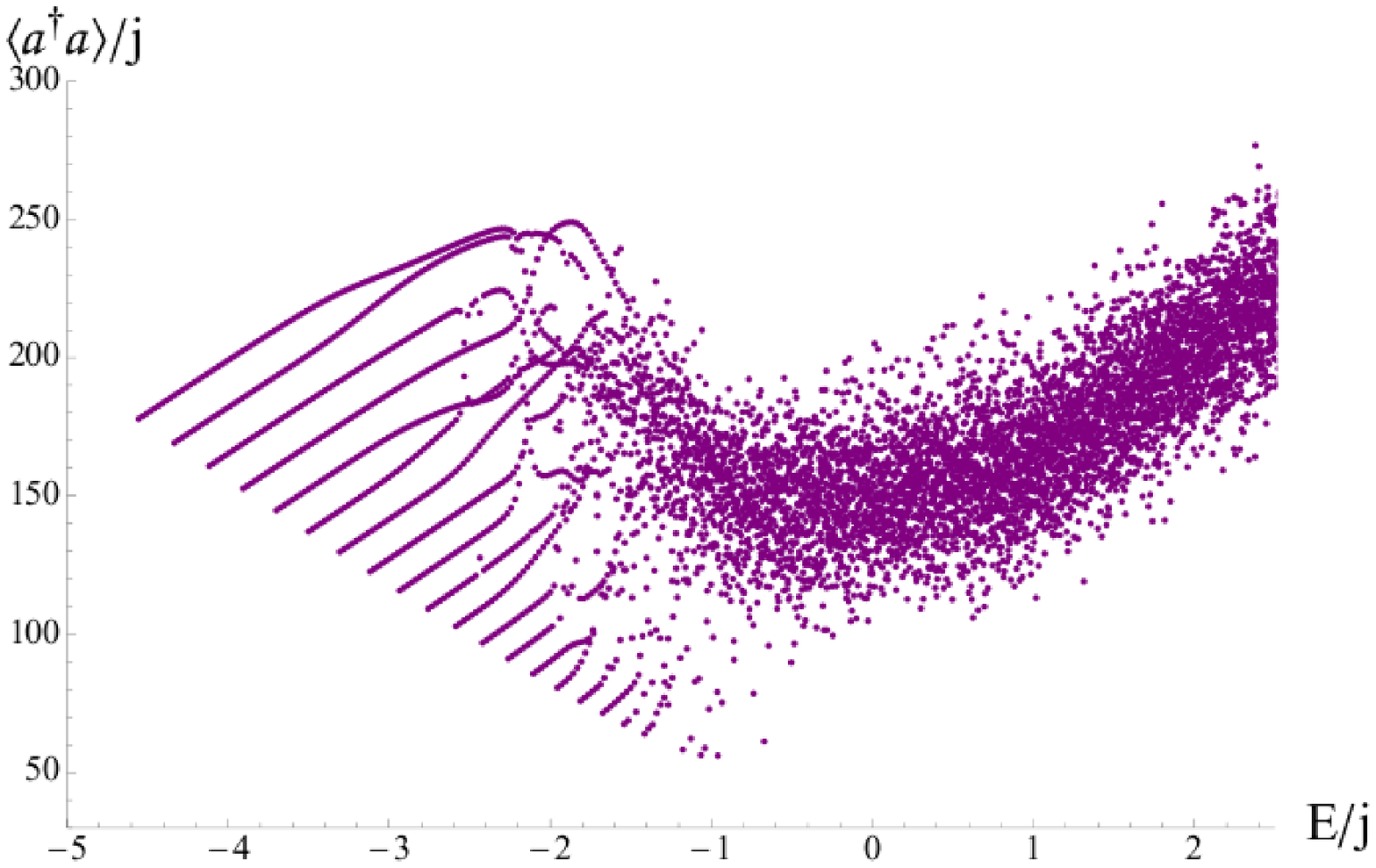} &\includegraphics[angle=0,width=0.5\textwidth]{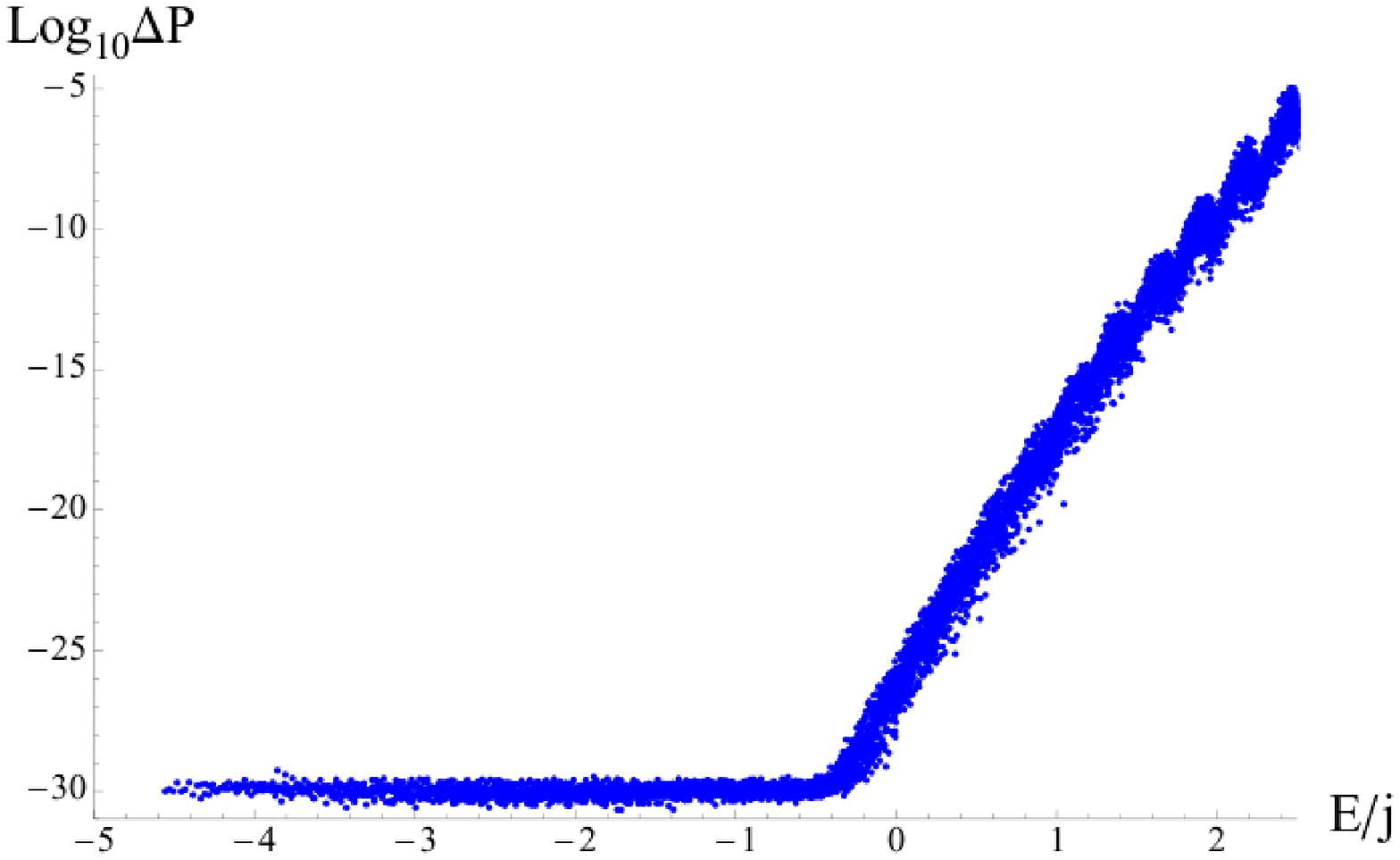} \\
\end{tabular}
\caption{Peres lattices for the Dicke model. Deep superradiant phase $\gamma=3\gamma_{c}$, with $N_{mnax}=300$. Peres operators: $J_{z}$ (a), $J_{x}^{2}$ (b) and $n$ (c). The numerical precision in the wave function for each individual state is shown in (d).}  
\label{fig5}
\end{figure} 

Next, we observe several representative couplings in the superradiant phase  where  the critical coupling is large than the critical one, $\gamma > \gamma_{c}$: $\gamma=1.5\gamma_{c}$ (figure \ref{fig3}),  $\gamma=2.0 \gamma_{c}$(figure \ref{fig4}), and, for the ultra strong coupling in the superradiant phase $\gamma=3\gamma_{c}$ (figure \ref{fig5}). Thanks to the lattices we can observe a pair of important things.

First, the $J_{z}$ Peres lattice exhibit two ESQPT. There are two changes in the slope of the lattice. One in $E/j=1$ and another in $E/j=-1$. In fact, the ESQPT can be seen too in figure \ref{fig1}. The $E/j=-1$ ESQPT is the one pointed by Perez-Fern\'andez et. al. \cite{Per11A}. From the $J_{z}$ lattice and following the results in \cite{Bran13}, it can be recognized that the two ESQPT are different in nature. The one in $E/j=1$ is \emph{static}, because as it can be observed from figure \ref{fig1} it corresponds to a saturation of the atomic space. For a given $\mathcal{N}$, there is a maximum value of excited atoms, the same $\mathcal{N}$. The \emph{static} ESQPT is independent of the coupling. On the other hand, the second one in $E/j=-1$ is a \emph{dynamical} ESQPT because it depends on the coupling. It is always associated with the state with $n=0$, $m=-j$, which in the superradiant phase is no longer the ground state. The strong atom-photon interaction builds up many states which combine a large number of photons and excited atoms with energies $E/j < -1$.

The second interesting feature is the break in the regularity. This is what one would expect from a non-integrable system. This is clear with every Peres operator. But also, there is a regular part for energies below $E/j=-1$. This regular region is small for weak couplings, but it grows with the coupling. Deep in the superradiant phase (figures \ref{fig4} and \ref{fig5}) it is clear that there is a transition between regularity and chaos. Besides, it seems the \emph{dynamical} ESQPT could be related with this transition. It would be necessary more work to determine if the onset of chaos is related to the ESQPT.   

\section{Conclusions}

We have employed the Peres lattices as a powerful tool to analyze the spectrum features of quantum systems. They helped us to determine the presence of two Excited State Quantum Phase Transitions and to distinguish between integrable and non-integrable regions. It gives us a way to define (qualitatively) integrability. The Dicke model is a good choice to study several interesting topics, with the great advantage to be a quantum non-integrable model which has well studied integrable limits. The lattices confirmed this but also reveal us the richness of the spectrum. A non-integrable system does not limit itself to a irregular behavior in the spectrum. In particular, for the Dicke model we can observe clearly the presence of two ESQPT, their nature, and a sort of transition between chaos and regularity. In a forthcoming work we present a join study involving the Peres method and quantitative analysis of quantum spectra, both numerical and analytical, which go beyond the scope of the present article \cite{Ler13}. 

We thank S. Lerma, P. Stransky and P. Cejnar for many useful and interesting conversations.This work was partially supported by CONACyT- M\'exico and DGAPA-UNAM.
 

\appendix
\section{Precision in the wave function}

We express the ground state wave function as
\begin{equation}
|\Psi(N_{max})\rangle=\sum\limits_{N=0}^{N_{max}} \sum\limits_{m=-j}^{j} C_{N,m} |N;j,m\rangle,
\end{equation} 
where $|N;j,m\rangle$ is the extended bosonic basis \cite{Chen0809}, \cite{Basta11}. Where $N$ is the eigenvalue of the $A^{\dagger}A$ operator ($A=a+\frac{2\gamma}{\sqrt{\mathcal{N}}\omega}J_{x}$), and $m$ are the eigenvalues of $J_{x}$. $C_{N,m}$ are the coefficients of the exact ground state wave function and $N_{max}$ is the value of the truncation or cutoff in the number of displaced excitations. The probability $P_{N}$ of having $N$ excitations in the ground state is
\begin{equation}
P_{N}=|\langle N|\Psi\rangle|^{2}=\sum_{m}|C_{N,m}|^{2}
\end{equation}
We define the precision in the calculated wave function as \cite{Basta13}:
\begin{equation}
\Delta P=\sum\limits_{m=-j}^j \left|C_{N_{max}+1,m}(N_{max}+1)\right|^2.
\end{equation}
By diagonalizing the Hamiltonian with several truncations, if $ \Delta P$ is smaller than certain tolerance we consider that the solution has converged, being $N_{max}$ the minimum value of the truncation necessary for obtaining the exact numerical solution. 

\section{Basis with well defined parity}

In order to solve numerically the Dicke Hamiltonian, we use the extended coherent bosonic basis with the right parity. It can be shown the action of the parity operator (Eq. \ref{parity}) over the coherent basis is:
\begin{equation}
\Pi|N;j,m\rangle=(-1)^{N}|N;j,-m\rangle
\end{equation}
So, the eigenvectors of the Hamiltonian in the limit $\omega_{0}\rightarrow 0$ and simultaneously of the parity operator are:
\begin{eqnarray}
 |N;j,m;\pm\rangle =\frac{1}{\sqrt{2(1+\delta_{m,0})}}\left(|N;j,m\rangle+(-1)^{N}|N;j,-m\rangle\right)
\end{eqnarray}
Using this basis, we can obtain energy levels which are deeper in the spectrum and also we can employ level statistics techniques to analyze it. 


\end{document}